\documentclass[jcp, reprint]{revtex4-1}

\usepackage{hyperref}
\usepackage{graphicx}
\usepackage{amstext}
\usepackage{amsmath}
\usepackage{amssymb} 

\def\be{\begin{equation}}
\def\ee{\end{equation}}
\def\bea{\begin{eqnarray}}
\def\eea{\end{eqnarray}}
\def\bal#1\eal{\begin{align*}#1\end{align*}}
\def\ba#1\ea{\begin{align}#1\end{align}}
\def\be{\begin{eqnarray}}
\def\ee{\end{eqnarray}}

\def\ud{\textrm{d}}

\newcommand{\bra}[1]{\langle #1|}
\newcommand{\ket}[1]{| #1\rangle}
\newcommand{\fig}[1]{Fig.~\ref{#1}}
\newcommand{\figs}[1]{Figs.~\ref{#1}}

\newcommand{\eq}[1]{Eq.~\eqref{#1}}

\newcommand{\vect}[1]{\mathbf{#1}}
\newcommand{\sv}[1]{\mathbf{#1}}
                         
\begin{document}
\title{Entanglement in the Born-Oppenheimer Approximation}
\author{Artur F. Izmaylov} %
\email{artur.izmaylov@utoronto.ca}
\affiliation{Department of Physical and Environmental Sciences,
  University of Toronto Scarborough, Toronto, Ontario, M1C 1A4,
  Canada; and Chemical Physics Theory Group, Department of Chemistry,
  University of Toronto, Toronto, Ontario, M5S 3H6, Canada}
\author{Ignacio Franco}
\email{ignacio.franco@rochester.edu}
\affiliation{Department of Chemistry and The Center for Coherence and Quantum Optics, University of Rochester, Rochester, New York 14627, USA}
\date{\today}
\begin{abstract}
The role of electron-nuclear entanglement on the validity of the Born-Oppenheimer (BO) approximation is investigated.   While nonadiabatic couplings generally lead to entanglement and to a failure of the BO approximation,  surprisingly the degree of electron-nuclear entanglement is found to be  uncorrelated with the degree of validity of the BO approximation. This is because while the degree of entanglement of BO states is determined by their deviation from the corresponding states in 
the crude BO approximation, the accuracy of the BO approximation is dictated, instead, by the deviation of the BO states from the exact electron-nuclear states.  In fact, in the context of a minimal avoided crossing model, extreme cases are identified where an adequate BO state is seen to be maximally entangled, and where the BO approximation fails but the associated BO state remains approximately unentangled. Further, the BO states are found  to not preserve the entanglement properties of the exact electron-nuclear eigenstates, and to be completely unentangled only in the limit in which the BO approximation becomes exact.
\end{abstract}

\maketitle
\section{Introduction}
The Born-Oppenheimer (BO) approximation forms the basis of our interpretation of chemical phenomena. As a consequence, considerable effort has been devoted to understand its scope, and to develop methods that allow us to think and model matter beyond its limits~\cite{Kapral:1999/jcp/8919, Tully:1990/jcp/1061,BenNun:2002tx, huo2011, Abedi:2010/prl/123002,Izmaylov:2011ev,Izmaylov:2013fe, Endicott:2014kl}. Surprisingly, however, an unexplored aspect of the BO approximation is its connection with entanglement~\cite{Schrodinger1935, EPR1935}, a basic quantum-mechanical correlation that is the essential resource for quantum information~\cite{Nielsen:2011}.  In addition to its interest at a fundamental level, understanding the role of entanglement in the BO picture is central in interpreting coherence phenomena in matter and in the development of methods to follow correlated  electron-nuclear dynamics.  Specifically,  to be able to capture all relevant  quantum correlations, approximate semiclassical or quantum descriptions of the electron-nuclear evolution of molecules~\cite{Kapral:1999/jcp/8919, Tully:1990/jcp/1061, Tully:2012/jcp/22A301, huo2011, subotnik2016} should preserve the entanglement character of the electron-nuclear states. Further,  the understanding of  coherence phenomena in molecules~\cite{martinez2007, fleming09, Engel:2007hb, Collini:2010fy,Ishizaki:2012kf,Pachon:2012fm, hannewald09, choi2010}, such as coherent spectroscopies, photoexcited dynamics and electron transfer events, requires a detailed understanding of the molecular events that lead to electronic decoherence through entanglement with the nuclear environment~\cite{schlosshauer, joos, kar2016, reducedpurities,  franco2013, franco2012}. 

To appreciate the non-trivial role of entanglement in the BO picture, consider the  exact wave-function of a pure electron-nuclear system in  a factorized form~\cite{Hunter1995, Abedi:2010/prl/123002, Abedi2012}
\be\label{eq:exfac}
\Psi(\vect{r},\vect{R}) =  \phi_{(e)}(\vect{r};\vect{R})\chi_{(e)}(\vect{R}),  
\ee
where 
\bea
\chi_{(e)}(\vect{R})=\left(\int d\vect{r} |\Psi(\vect{r},\vect{R})|^2\right)^{1/2} 
\eea
is the nuclear wave-function (\protect{$\int \ud\vect{R}  |\chi_{(e)}(\vect{R})|^2 = 1$}),  
and 
\bea
\phi_{(e)}(\vect{r}; \vect{R}) = \Psi(\vect{r},\vect{R})/\chi_{(e)}(\vect{R})
\eea
is the conditional probability amplitude of finding electrons at $\vect{r}$ given that the nuclear configuration 
is  $\vect{R}$ (\protect{$\int \ud\vect{r} |\phi_{(e)}(\vect{r};\vect{R})|^2 =1$}). This exact decomposition represents  an entangled electron-nuclear state because of the dependency of the electronic conditional probability amplitude $\phi_{(e)}(\vect{r}; \vect{R})$ on $\vect{R}$. By contrast, in the \emph{crude} Born-Oppenheimer (CBO) approach, the electron-nuclear wave-function is approximated as
\bea\label{eq:CBO}
\Psi(\vect{r},\vect{R}) \approx  \phi(\vect{r};\vect{R}_0)\tilde{\chi}(\vect{R}), 
\eea
where $\phi(\vect{r}; \vect{R}_0)$ is an eigenfunction of the electronic Hamiltonian {$H_e(\vect{r};\vect{R})$}
for a particular nuclear configuration {$\vect{R}=\vect{R}_0$, i.e.}
\bea
\hat H_e(\vect{r};\vect{R}_0) \phi(\vect{r};\vect{R}_0) = E(\vect{R}_0) \phi(\vect{r};\vect{R}_0),
\eea
and $\tilde{\chi}(\vect{R})$ is the nuclear counterpart. 
Considering that a CBO state is a {separable} product between a nuclear state and an electronic state [\eq{eq:CBO}],
the electron-nuclear state is clearly unentangled.
The BO states are intermediate between these two limiting situations
\bea\label{eq:BO}
\Psi(\vect{r},\vect{R}) \approx  \Psi_{\rm BO}(\vect{r},\vect{R})=\phi(\vect{r};\vect{R})\chi(\vect{R}),
\eea
where the electronic function $\phi(\vect{r}; \vect{R})$ is obtained as an eigen-function of the electronic 
Hamiltonian for all nuclear configurations
\bea
\hat H_e(\vect{r};\vect{R}) \phi(\vect{r};\vect{R}) = E(\vect{R}) \phi(\vect{r};\vect{R}).
\eea
Thus, $\phi(\vect{r}; \vect{R})$ \emph{is} allowed an $\vect{R}$ dependence, 
suggesting that BO states are entangled.
 However, such a dependency is restricted so that nuclear motion proceeds without changes in the quantum state of the electron cloud, suggesting that the nuclear and electronic dynamics are somewhat 
 less correlated.  The main questions are: How entangled are BO states? Does the BO approximation preserves the entanglement character of the exact states? Is there a relation between the degree of entanglement {of electron-nuclear states} and  the validity of the BO approximation?

Here, we address these questions and clarify the role of electron-nuclear entanglement in the BO approximation. This is done through formal considerations and model computations in a two-state one-dimensional system  with an avoided crossing.  Specifically, we show that BO states are generally entangled except in the limit in which the BO approximation becomes exact. Interestingly, while non-adiabatic couplings can lead to entanglement and to a failure of the BO approximation, entanglement does not necessarily lead to a significant failure of the BO  approximation.
  
The structure of this paper is as follows: Section \ref{sec:theorA} introduces purity as a measure of entanglement in the context of electron-nuclear systems. This purity measure is used in Sec.~\ref{sec:theorB} to analyze  the general entangled character of exact and BO states, and the unentangled limit in which the BO approximation becomes formally exact. Section~\ref{sec:theorC} introduces a one-dimensional two-state model with an avoided crossing that is used to  illustrate numerically the correlation, or lack thereof, between the accuracy of the BO approximation and the degree of entanglement. In Sec.~\ref{sec:concl} we summarize our main findings and discuss their implication in the interpretation of coherence phenomena.
 
 \section{Purity as a measure of entanglement}
  \label{sec:theorA}
  
As a measure of entanglement between electrons and nuclei we can employ the purity of either the electronic 
\be
P_e  = \textrm{Tr}\{\hat{\rho}_e^2\},
\ee
or nuclear subsystem 
\be
P_N  = \textrm{Tr}\{\hat{\rho}_N^2\},
\ee
where $\hat{\rho}_e = \textrm{Tr}_N \{ \hat{\rho} \}$  ($\hat{\rho}_N = \textrm{Tr}_e\{ \hat{\rho} \}$) is the electronic (nuclear) {reduced} density matrix obtained by tracing out the nuclear (electronic) 
degrees of freedom out of the density matrix of the full system $\hat{\rho} = \ket{\Psi}\bra{\Psi}$. 
For unentangled electron-nuclear systems, $P_e= P_N=1$, while entanglement leads to a 
non-idempotency of the reduced density matrix and thus $P_e$ and $P_N$ values 
lower than $1$. Such entanglement is a basic source of electronic (or nuclear) decoherence as it leads to a mixed density matrix for the electronic (or nuclear) subsystem. 

As a consequence of the Schmidt theorem~\cite{Schmidt1907, Nielsen:2011}, for pure electron-nuclear systems the electronic and nuclear purity actually coincide, i.e. $P_e = P_N=P$. 
The Schmidt theorem can be readily verified for the  general electron-nuclear state in \eq{eq:exfac}. The nuclear reduced density matrix is given by
\be
\label{eq:ndm}
\rho_N(\vect{R},\vect{R'}) = \chi_{(e)}(\vect{R})\chi_{(e)}^{\star}(\vect{R'}) \int \ud\vect{r}  \,
\phi_{(e)}(\vect{r}; \vect{R})\phi_{(e)}^\star(\vect{r}; \vect{R'}),
\ee
where the integral is not a normalization integral since $\vect{R}$ and $\vect{R'}$ can have different values. 
The corresponding electronic density matrix is 
\be
\label{eq:edm}
\rho_e(\vect{r}, \vect{r'}) = \int \ud\vect{R}\, |\chi_{(e)}(\vect{R})|^2 \phi_{(e)}(\vect{r};\vect{R})  \phi_{(e)}^\star(\vect{r}';\vect{R}). 
\ee
The purity of the electronic and nuclear state coincide since
\be
\begin{split}
\textrm{Tr}\{\hat{\rho}_e^2\}  & = \int \ud\vect{r}\ud\vect{r'} \rho_e(\vect{r}, \vect{r'}) \rho_e(\vect{r'}, \vect{r}) \\
& = \int \ud\vect{r}\ud\vect{r'}\ud\vect{R}\ud\vect{R'} | \chi_{(e)}(\vect{R})|^2 | \chi_{(e)}(\vect{R'})|^2 \times \\ 
& \phi_{(e)}^\star(\vect{r};\vect{R}) 
\phi_{(e)}(\vect{r}';\vect{R})\phi_{(e)}^\star(\vect{r}';\vect{R'}) \phi_{(e)}(\vect{r};\vect{R'})\\ 
 & =         \int \ud\vect{R}\ud\vect{R'} \rho_N(\vect{R}, \vect{R'}) \rho_N(\vect{R'}, \vect{R}) \\
 & = \textrm{Tr}\{ \hat{\rho}_N^2 \}= P.
\end{split}
\ee
 Naturally, this theorem also applies to BO electron-nuclear states. Therefore, without loss of generality, to quantify entanglement in the exact and BO case we can focus on the purity of the nuclear subsystem. As it turns out, this choice is particularly convenient because of the inherent asymmetry of the BO state.  

 \section{Entanglement of electron-nuclear states}
 \label{sec:theorB}
 
To illustrate the entanglement in terms of purity for the exact [\eq{eq:exfac}] and BO [{\eq{eq:BO}}]
electron-nuclear states it is instructive to consider an expansion of their 
electronic components in the CBO basis $\{\phi_i(\vect{r}; \vect{R}_0)\}$. Since the algebra
involved in our consideration is exactly the same for the exact electronic conditional probability and  
BO electronic state we will derive the purity expression only for the BO case. 
The electronic BO function can be written as
\bea\label{eq:CBOexp}
\phi(\vect{r}; \vect{R}) = \sum_i\phi_i(\vect{r}; \vect{R}_0) C_i(\vect{R}),
\eea
which makes the BO electron-nuclear function  
\be
\begin{split}
\Psi_{\rm BO}(\vect{r}, \vect{R}) &= \sum_i\phi_i(\vect{r}; \vect{R}_0) C_i(\vect{R}) \chi(\vect{R}) \\
&= \sum_i\phi_i(\vect{r}; \vect{R}_0) \tilde{\chi}_i(\vect{R}).
\end{split}
\ee
Here 
\be
\int \ud\vect{R}  |\tilde{\chi}_i(\vect{R})|^2  = \int \ud\vect{R}  |\chi(\vect{R})|^2 |C_i(\vect{R})|^2 \le 1,
\ee
which is a consequence of the positivity of the absolute squares and the unit upper boundary 
for the $|C_i(\vect{R})|^2$ function. The equality is only possible for the case when there is only one term in the CBO expansion [\eq{eq:CBOexp}].

Using the CBO basis, the nuclear density expands as
 \be
 \begin{split}
\label{TSe1}
\rho_N(\vect{R},\vect{R'}) & =
 \sum_{ij}\tilde{\chi}_i(\vect{R})\tilde{\chi}_j^\star(\vect{R'}) 
 \bra{\phi_j(\vect{R}_0)}\phi_i(\vect{R}_0)\rangle \\
&= \sum_{i}\tilde{\chi}_i(\vect{R})\tilde{\chi}_i^\star(\vect{R'}),
\end{split}
\ee
where we have taken into account the orthogonality of the CBO states.
This form is convenient for illustrating the fact that the nuclear density matrix represents a mixed state
due to entanglement: 
\be
\begin{split}
P=\textrm{Tr}\{ \hat{\rho}_N^2 \} &= \int \ud\vect{R} \ud\vect{R'} \rho_N(\vect{R},\vect{R'})\rho_N(\vect{R'},\vect{R}) \\
&= \sum_{ij} \bra{\tilde{\chi}_j}\tilde{\chi}_i\rangle  \bra{\tilde{\chi}_i}\tilde{\chi}_j\rangle.
\end{split}
\ee
Substituting 
$S_{ij}=\bra{\tilde{\chi}_i}\tilde{\chi}_j\rangle$
we have
\be
\label{eq:ovlp}
P=  \sum_{ij} |S_{ij}|^2 
\le \sum_{ij} S_{ii} S_{jj}= 1,
\ee
where we  have taken into account the Schwarz inequality and the 
normalization condition $\sum_i S_{ii} =1$. {The equality 
in \eq{eq:ovlp} is only possible for two special cases: when there is only 
a single term in the CBO expansion or when all nuclear components
$\tilde{\chi}_i$ are equal to each other up to constant multiplicative factors.
Thus, \eq{eq:ovlp} clearly shows that for general electron-nuclear states,
exact or BO, are generally entangled, as expected.

A complementary perspective on the origin of entanglement in electron-nuclear states 
can be gleaned from the purity of the BO state without performing the CBO expansion
\be
\label{eq:purN}
P = \int \ud\vect{R}\ud\vect{R'} | \chi(\vect{R})|^2 | \chi(\vect{R'})|^2  |\bra{\phi(\vect{R})}\phi(\vect{R'})\rangle|^2. 
\ee
The electronic part $|\bra{\phi(\vect{R})}\phi(\vect{R'})\rangle|^2$ can be bound from 
above by the Schwartz inequality 
\bea\label{eq:elSchw}
|\bra{\phi(\vect{R})}\phi(\vect{R'})\rangle|^2 \le \bra{\phi(\vect{R})}\phi(\vect{R})\rangle
\bra{\phi(\vect{R'})}\phi(\vect{R'})\rangle = 1.
\eea
As in \eq{eq:ovlp}, for general electronic wavefunctions the equality in \eq{eq:elSchw} is not relevant as any nuclear dependence in 
$|\bra{\phi(\vect{R})}\phi(\vect{R'})\rangle|^2$ leads to values lower than 1. Taking into account the normalization of $\chi(\vect{R})$ and that $|\chi(\vect{R})|^2>1$,   it follows that any nuclear 
dependence in $|\bra{\phi(\vect{R})}\phi(\vect{R'})\rangle|^2$ also results in $P<1$. Naturally, this result is consistent with the CBO expansion because
a nuclear dependence of the electronic wave-function leads to multiple terms in the 
CBO expansion.  
To elucidate this dependence let us consider the expansion of the BO electronic wavefunction $\phi(\vect{r}; \vect{R'})= \phi(\vect{r}; \vect{R}+\vect{a})$  in \eq{eq:ndm} around $\vect{R}'=\vect{R}$ ($\vect{a}=0$):
\be
\label{eq:phishifted}
\begin{split}
\phi(\vect{r}; \vect{R}+\vect{a}) & = \exp\left(\frac{i}{\hbar}\vect{a}\cdot\hat{\vect{P}}\right) \phi(\vect{r}; \vect{R})\\
& = \sum_{n=0}^\infty\frac{1}{n!}\left(\frac{i}{\hbar}\vect{a}\cdot\hat{\vect{P}}\right)^n \phi(\vect{r}; \vect{R}),
\end{split}
\ee
where $\hat{\vect{P}} =  -i\hbar\partial_\vect{R}$ is the total nuclear momentum operator. 
Inserting \eq{eq:phishifted} into \eq{eq:ndm} yields
\begin{multline}
\label{eq:nrdmtmp}
\rho_N(\vect{R},\vect{R}+ \vect{a})   = \chi(\vect{R})\chi^\star(\vect{R}+ \vect{a})( 1 +\\
\sum_{n=1}^\infty\frac{1}{n!} \bra{\phi (\vect{R})} \left(\frac{i}{\hbar}\vect{a}\cdot\hat{\vect{P}}\right)^n\ket{\phi(\vect{R})}).
\end{multline}
The first term corresponds to the pure (idempotent) nuclear density matrix. 
Any entanglement is introduced by the second term governed by the derivatives of the electronic wave-functions with respect to the nuclear coordinates. 
The BO approximation assumes a weak dependence of the electronic wave-functions on the 
nuclear configuration. In the limit where the BO approximation is \emph{exact} all
derivatives in \eq{eq:nrdmtmp} should be zero. In this limit, the electron-nuclear states become 
unentangled, and the expansion of the electron-nuclear state in the CBO basis consists of 
only one term [cf. \eq{eq:CBOexp}]. 
However, note that even a mild dependence of the electronic states on the nuclear coordinates can lead to appreciable entanglement. This effect is particularly important when the nuclear state is 
highly delocalized in space such that the terms $\chi(\vect{R})\chi^\star(\vect{R}+ \vect{a}) \bra{\phi(\vect{R})} \left(\frac{i}{\hbar}\vect{a}\cdot\hat{\vect{P}}\right)^n\ket{\phi(\vect{R})}$ are appreciable even for large $n$'s.
The delocalization of the nuclear wave-function makes both $\chi(\vect{R})$ and  $\chi(\vect{R}+ \vect{a})$ 
appreciable even for large $||\vect{a}||$.  In turn, a large $||\vect{a}||$ enhances the whole term due to its 
$n^{\rm th}$ power even for small derivatives of the electronic wave-function.  
For this reason, as discussed in Sec.~\ref{sec:num}, for nuclear states with a strong degree of spatial delocalization the adequacy of the BO approximation is not necessarily correlated with the degree of entanglement of the states.}
  
\section{Entanglement in an avoided crossing model}
 \label{sec:theorC}
 
We exemplify the relation between entanglement and the validity of the BO approximation on a minimal model 
for {an} avoided crossing (AC) problem (or one-dimensional spin-boson model)~\cite{Book/Nitzan:2006}.  
The AC model is one of the simplest cases where breakdown of the BO approximation
can be {modeled} easily~\cite{Landry:2011ej,Gherib:2016ch}.

\subsection{Theory and Model}

\paragraph{Model Hamiltonian:}
We introduce two diabatic states, $\ket{\varphi_1}$ and $\ket{\varphi_2}$, 
which will represent the complete set of CBO basis functions $\{\phi_i(\vect{r};\vect{R}_0)\}$
\footnote{This equivalence is true up to a constant unitary transformation diagonalizing a matrix of the electronic Hamiltonian within the two-state subspace at  the $\vect{R_0}$ configuration.},  
and whose explicit electronic coordinate dependence will not be of importance. 
Presenting the total wave-function as 
\bea\label{eq:exact}
\ket{\Psi} = \ket{\tilde{\chi}_1}\ket{\varphi_1} +\ket{\tilde{\chi}_2}\ket{\varphi_2},
\eea
we project the total time-independent Schr\"odinger equation (TISE) onto  
the electronic states $\{\ket{\varphi_i}\}_{i=1,2}$
\bea\label{Exact}
   \hat H \begin{pmatrix} 
    \ket{\tilde{\chi}_1}  \\
    \ket{\tilde{\chi}_2}
  \end{pmatrix}
  = E \begin{pmatrix} 
    \ket{\tilde{\chi}_1}  \\
    \ket{\tilde{\chi}_2}
  \end{pmatrix},
\eea
where
\begin{equation}
  \label{eq:H_sb}
  \hat H = {\hat T}  {\mathbf 1}_2 + 
  \begin{pmatrix} 
    V_{11} & V_{12} \\
    V_{12} & V_{22}
  \end{pmatrix},
\end{equation}
$\hat T = -\frac{1}{2} \partial_x^2 $ is the nuclear kinetic
energy operator (the units are chosen such that $\hbar=m=1$ and, for simplicity, the nuclear subspace contains 
only one coordinate $\sv{R}=x$), and ${\mathbf 1}_2$ is a $2\times 2$ unit matrix.  
The diabatic potentials $V_{11}$ and $ V_{22}$ are identical 1D
parabolas shifted in the $x$-direction by $a$ {and} in energy by $\Delta$, {i.e.}
\begin{align}
  \label{eq:diab-me-11}
  V_{11} = {} & \frac{\omega^2x^2}{2},\\
  \label{eq:diab-me-22}
  V_{22} = {} & \frac{\omega^2}{2}(x - a)^2 + \Delta.
\end{align}
To have an avoided crossing in the adiabatic representation $V_{11}$ and $V_{22}$ 
are coupled by a constant  potential $V_{12}=c$.

Switching to the adiabatic representation for the 1D AC Hamiltonian in 
\eq{eq:H_sb} is done by diagonalizing the potential
matrix using a unitary transformation 
\begin{equation}
  U = 
  \label{eq:Umat}
  \begin{pmatrix}
    \cos\theta & \sin\theta \\
    -\sin\theta & \cos\theta
  \end{pmatrix},
\end{equation}
where $\theta=\theta(x)$ is a mixing angle {in the superposition} between the diabatic electronic
states states $\ket{\varphi_1}$ and $\ket{\varphi_2}$, and is given by
\begin{equation}
  \label{eq:theta}
  \theta = \frac{1}{2}\arctan \dfrac{2\,V_{12}}{V_{11} - V_{22}} =
  \frac{1}{2}\arctan \dfrac{\gamma}{x - b}.
\end{equation}
Here, $b = \Delta/(\omega^2 a)$ is the $x$-coordinate of the crossing
point, and 
\be
\label{eq:gamma}
\gamma = {2c}/{(\omega^2a)}
\ee
 is a coupling strength between the diabatic states. 

The transformation $U(\theta)$ defines the BO electronic states 
\begin{eqnarray}\label{eq:adi1}
  \ket{\phi_1(x)} & = & \phantom{-}\cos\theta\,\ket{\varphi_1} +
  \sin\theta\,\ket{\varphi_2} \\ \label{eq:adi2}
  \ket{\phi_2(x)} & = & -\sin\theta\,\ket{\varphi_1} +
  \cos\theta\,\ket{\varphi_2}
\end{eqnarray}
and gives rise to the 1D AC
Hamiltonian in the adiabatic representation $\hat H_\text{adi}  = U
{\hat H} U^\dagger$,
\begin{equation}
  \label{eq:adiab}
   \hat H_\text{adi}  =   
  \begin{pmatrix}
    \hat T + \hat\tau_{11}& \hat\tau_{12} \\
    \hat\tau_{21} & \hat T +\hat\tau_{22}
  \end{pmatrix} +
  \begin{pmatrix}
    W_{-} & 0 \\
    0 & W_{+}
  \end{pmatrix},
\end{equation}
where
\begin{align}
  \label{eq:Wmin}
  W_{\pm} = & {} \dfrac{1}{2}\left(V_{11} + V_{22}\right) \pm
  \dfrac{1}{2}\sqrt{\left(V_{11} - V_{22}\right)^2 + 4 V_{12}^2}
\end{align}
are the adiabatic potentials and $\hat\tau_{ij}=-\left\langle
  \phi_i(x) | \partial_x \phi_j(x)\right\rangle\partial_x
- \frac{1}{2} \left\langle \phi_i(x) | \partial_x^2
  \phi_j(x)\right\rangle$ are the nonadiabatic couplings (NACs). For this model, {the}
 NACs can be expressed as  
 \bea
 \hat\tau_{11} &=& \hat\tau_{22} = \frac{1}{2} [\partial_x \theta(x)]^2, \\
 \hat\tau_{21} &=& -\hat\tau_{12} = - \partial_x\theta(x)\partial_x-\frac{1}{2} \partial_x^2 \theta(x).
 \eea
The BO approximation neglects all nonadiabatic terms $\hat\tau_{ij}$ and 
formulates the nuclear TISE as
\bea
[\hat T + W_{\pm}(x) ] \chi(x) = E_{BO}\chi(x).
\eea
The adequacy of the BO approximation in this model depends on the  NAC element 
\be
\begin{split}
\langle\phi_2(x) | \partial_x \phi_1(x)\rangle =& \partial_x\theta(x)  \\ \label{eq:d21}
=&\frac{\gamma}{4\gamma^2+(x-b)^2}.
\end{split}
\ee
The maximum of $\partial_x\theta(x)$ is at the crossing point $x=b$ and has a simple 
dependence on model parameters 
\be\label{eq:thmax}
(\partial_x\theta)_{\rm max} = \frac{1}{4\gamma}.
\ee
Although it may seem that  \eq{eq:thmax} provides a straightforward way to predict 
 the  failure or success of the BO approximation, to get an accurate assessment 
one also needs to consider the nuclear density at the vicinity 
of the crossing point $b$.  This is because  $\langle\phi_2(x) | \partial_x \phi_1(x)\rangle$
is  part of the nuclear kinetic energy operator in \eq{eq:adiab} and, therefore, without 
non-negligible nuclear density a large NAC value will not have a  significant effect. 

\paragraph{Purity:}
As for entanglement measured in terms of the purity, \eq{eq:ovlp} for this two-state case can 
be expressed as 
\be \label{eq:pur2s}
\begin{split}{\rm Tr}[\hat\rho_N^2] =& (S_{11}+S_{22})^2-2(S_{11}S_{22}-S_{12}^2) \\
=& 1-2(S_{11}S_{22}-S_{12}^2). 
\end{split}
\ee
This shows that the loss of purity comes from the interplay between diagonal and off-diagonal 
nuclear overlap matrix elements $S_{ij}$.  For exact and BO states, we  will refer to nuclear states with $S_{ii}\gg S_{jj}$  as \emph{localized} and those with   $S_{11}\approx S_{22}\approx1/2$ as \emph{delocalized}.  Note that due to the Schwarz inequality $ S_{11} S_{22}\ge S_{12}^2$ and the normalization condition
 $S_{11}+S_{22}=1$, the localization condition $S_{ii}\gg S_{jj}$ always leads to vanishing $S_{12}^2$, whereas  
the delocalization condition $S_{11}\approx  S_{22}\approx1/2$ does not require $ S_{12}$ to be small.

\begin{figure}
  \centering
  \includegraphics[width=0.49\textwidth]{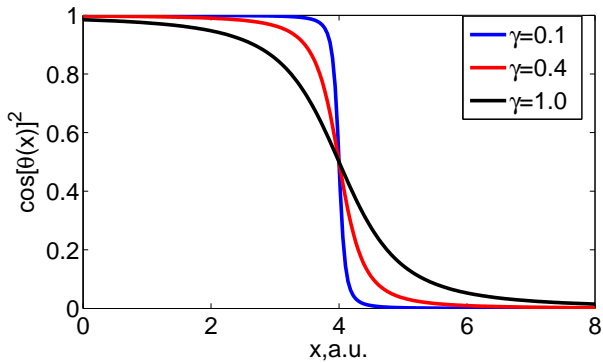}
  \caption{The partitioning function $\cos[\theta(x)]^2$ for different $\gamma$'s and the 
  crossing point for two parabolas $b=4$ a.u.}
  \label{fig:part}
\end{figure}

\paragraph{BO wave-function:} To analyze entanglement in the BO wave-function,
without loss of generality~\footnote{All subsequent 
derivations for the $S_{12}$ quantity can be repeated with the excited 
electronic state arriving at the same outcome.} 
we focus on the ground electronic state  wave-function expressed 
in the diabatic basis
\be
\begin{split}
\langle x\ket{\Psi_{BO}} =& \langle x\ket{\chi}\ket{\phi_1(x)}\\
=& \langle x\ket{\chi}(C_1(x)\ket{\varphi_1}+C_2(x)\ket{\varphi_2}) \\
=& \langle x\ket{\tilde{\chi}_1}\ket{\varphi_1} +\langle x\ket{\tilde{\chi}_2} \ket{\varphi_2},
\end{split}
\ee
where $\bra{x}\tilde{\chi}_1\rangle = \chi(x)\cos{[\theta(x)]}$ and 
$\bra{x}\tilde{\chi}_2\rangle = \chi(x)\sin{[\theta(x)]}$ are the nuclear components in 
the diabatic basis. 
The terms $\cos[\theta(x)]^2$ and $\sin[\theta(x)]^2$, depicted in \fig{fig:part}, can be thought as partitioning functions that split the nuclear 
BO probability density $|\chi|^2$ into the diabatic components $|\tilde{\chi}_1|^2$ and $|\tilde{\chi}_2|^2$.
As exemplified in \fig{fig:part}, these complementary partitioning functions go from 0 to 1 around $x=b$ in a characteristic length proportional to $\gamma$. This $\gamma$-dependence arises because 
$\partial_x \cos[\theta(x)]^2\sim\partial_x \sin[\theta(x)]^2\sim\partial_x \theta(x)$
and \eq{eq:d21}. 

Using these partitioning functions, 
the $S_{12}^2$ part of \eq{eq:pur2s} can be expressed as  
\be\label{eq:S12}
\begin{split}
S_{12}^2 
 =& \left(\int dx \chi^2(x) \cos[\theta(x)]\sin[\theta(x)]\right)^2 \\
 =& \frac{1}{4}\left(\int dx \chi^2(x) \sin\left[\arctan \left(\dfrac{\gamma}{x - b}\right)\right]\right)^2 \\
  =& \frac{1}{4}\left(\int dx \frac{\chi^2(x)} {\sqrt{\gamma^2+(x - b)^2}}\right)^2.
 \end{split}
 \ee
 Therefore, $S_{12}$ will be large if the BO nuclear probability density $\chi^2(x)$ 
 is high at the intersection point $x=b$.
Also, $S_{12}^2$ can be bound from above using the Schwarz inequality 
\be\label{eq:s12g}
\begin{split}
S_{12}^2 \le& \frac{1}{4}\int dx \chi^4(x)\int \frac{dx} {\gamma^2+(x - b)^2} \\
=& \gamma\frac{\pi}{4}\int dx \chi^4(x). 
\end{split}
\ee
Hence, $S_{12}^2\sim\gamma$, which allows us to simplify the purity in 
the limiting case of divergent NACs (recall \eq{eq:thmax})
\bea
\lim_{\gamma\rightarrow 0} P_N = 1-2S_{11}S_{22} \approx\left\{
\begin{array}{l}\label{eq:loc}
1,\,{\rm if}\, S_{ii}\gg S_{jj} \\ \label{eq:deloc}
1/2, \, {\rm if}~ S_{ii}\approx 1/2.
\end{array}
\right.
\eea

In turn, when $\gamma$ is appreciable,
$S_{12}^2$ can become comparable with $S_{11}S_{22}$ and this leads to an increased purity 
up to $P_N\approx1$. In the limit of $\gamma\rightarrow\infty$ the BO approximation is exact and the purity goes to 1. One of the 
simplest ways to see this is to consider the $a\rightarrow 0$ approach to the $\gamma\rightarrow\infty$
limit. If $a\rightarrow 0$, the two parabolas will always be parallel to each other and a unitary
transformation diagonalizing the potential matrix [\eq{eq:H_sb}] in one nuclear configuration
will diagonalize it for all other configurations. Therefore the CBO and BO states will be identical
which is enough for the purity to be 1 [see \eq{eq:ovlp}]. 
Other approaches to the $\gamma\rightarrow\infty$
limit ($\omega\rightarrow 0$ and $c\rightarrow\infty$)  give the same result.  
Also, in the $\gamma\rightarrow \infty$ limit the exact and BO states coincide because 
nonadiabatic couplings are zero. Therefore we will not focus on large $\gamma$'s in the numerical examples presented below.
 
\subsection{Numerical Examples}
\label{sec:num} 

To quantitatively investigate the correlation between the adequacy of the BO approximation and the degree of entanglement, we consider  three model cases defined by the parameters in Table \ref{tab:models}. The PES associated with each of the models are shown in Figs.~\ref{fig:m1}a-\ref{fig:m3}a. In each case, entanglement is quantified through the purity, while the adequacy of the BO approximation is assessed by examining the 
difference of the exact total energy $E$ [\eq{Exact}] with that obtained in the BO approximation $E_{BO}$, as well as 
the magnitude of the overlap between the corresponding electron-nuclear wave-functions 
$|\bra{\Psi}\Psi_{BO}\rangle|$. 

\subsubsection {Entanglement  in stationary states}

 \begin{table}[!h]
  \caption{Parameters of the three model two-level systems with Hamiltonian \eq{eq:H_sb}. In all models, $\omega=1$ and $c=\omega/5$ a.u.} 
  \label{tab:models}
  \centering
  \begin{ruledtabular}
    \begin{tabular}{@{}cccc@{}}
      \multicolumn{1}{c}{Model} & $a$ & \multicolumn{1}{c}{$\Delta$}  & $\gamma$ \\ \hline
      1 & 4 & 0 & 0.1 \\[1ex]
      2 & 4 & $1.5~\omega$ &  0.1 \\ [1ex]
      3 & 1 & $4~\omega$ & 0.4 \\ [1ex] 
    \end{tabular}
  \end{ruledtabular}
\end{table}

\begin{figure*}
  \centering
  \includegraphics[width=\textwidth]{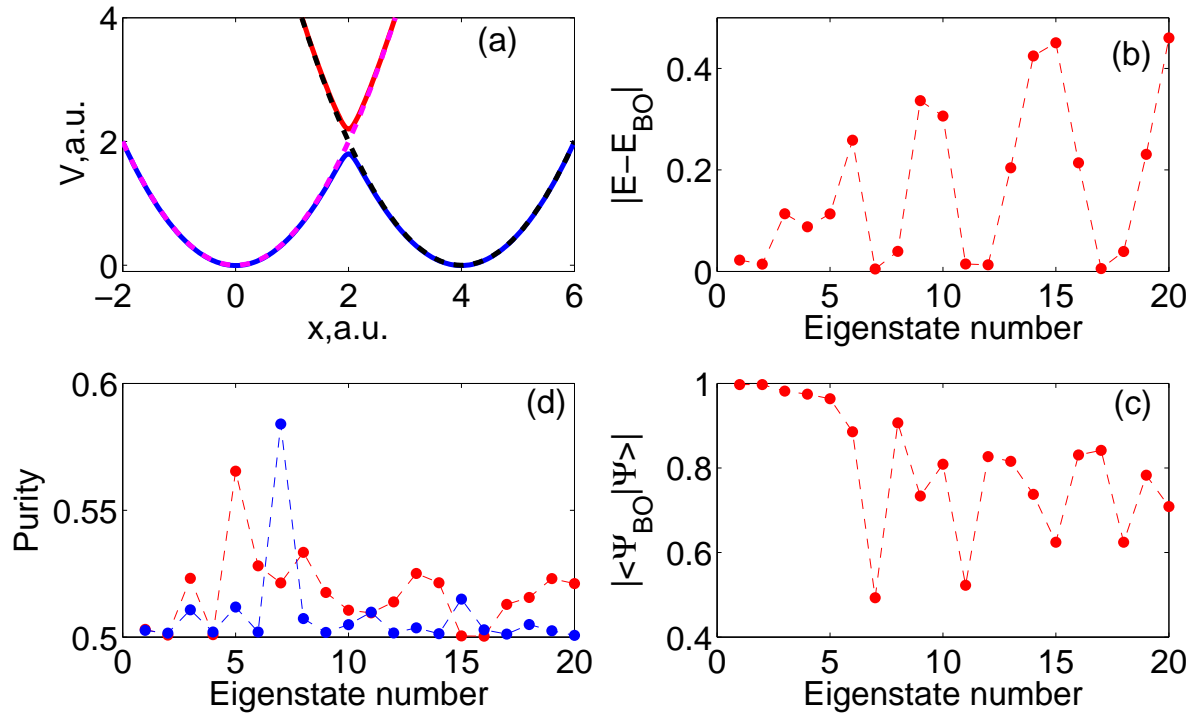}
  \caption{Model 1: (a) Adiabats (solid) and diabats (dashed); 
  (b) Absolute energy differences between exact and BO eigen-states;
  (c) Absolute overlaps between exact and BO eigen-functions; (d) Purities of the BO (blue) and exact (red) eigen-states.}
  \label{fig:m1}
\end{figure*}
\paragraph{Model 1:} Corresponds to a case in which the ground PES has two degenerate minima, while the excited PES has a single minimum located at the mid-point between the ground-state minima, see \fig{fig:m1}a. The first few states in this model are well 
described by the BO approximation (\fig{fig:m1}b and c) 
because their nuclear wave-functions do not have large probability density
in the  vicinity of the NAC function maximum [\eq{eq:thmax}].
The first state that has a substantial deviation from the exact wave-function according to the 
overlap criterion is the  7$^{th}$ BO state. The reason for the large discrepancy 
is that the  7$^{th}$ BO state is the ground vibrational 
state on the excited electronic BO-PES. Thus, it has large nuclear probability density in the vicinity of the 
 NAC maximum. Higher energy states in the BO approximation 
correspond to either the excited or the ground electronic states. Generally,
the overlap with the exact wave-function is better for the BO states of the ground electronic state.
Since the ordering of the eigen-states is done based on their energies, alternation
of states from the ground and excited BO-PES creates oscillations in absolute overlaps 
in \fig{fig:m1}c. 

For most states considered, the purity is close to $1/2$ both in the exact and BO 
 treatments (\fig{fig:m1}d). That is, the BO states approximately preserve the entangled character of the exact states. This takes place for both methods because 
nuclear overlaps $S_{12}$ are small and the  nuclear component of the eigenstates is  delocalized
($S_{11}\approx S_{22}\approx 1/2$). 
To understand the delocalization in the exact wave-functions one can use  first order perturbation theory 
to estimate {the relative contributions} of the diabatic vibrational states to the nuclear component of the exact wave-function: 
The low coupling ($c$) between the diabatic vibrational states is overpowered by 
the vibrational level alignment of two parabolas ({as} $\Delta=0$). Degenerate perturbation 
theory yields equal contributions of the two diabatic degenerate states to the eigenstates.  
In this case, the $S_{12}$ elements are Franck-Condon overlaps between energetically aligned 
diabatic vibrational states and they are small as a result of a relatively large spatial shift $a=4$.
In turn, the BO states are generally delocalized due to the symmetry of BO-PESs $W_{\pm}$. The low magnitude for the overlaps $S_{12}$ is a result of a small $\gamma$ [\eq{eq:s12g}]. 
The only appreciable increase in purity across the BO states 
can be seen for the 7$^{th}$ state, because of the highest localization of nuclear probability density 
at the crossing region in this state and as a result an increased overlap $S_{12}$ 
[see Eqs.~\eqref{eq:S12} and \eqref{eq:pur2s}]. 

In this model, the BO eigenstates provide a clear example of a system in which the BO approximation is appropriate (albeit not exact) but where the (exact and BO) electron-nuclear states are maximally entangled.

\begin{figure*}
  \centering
  \includegraphics[width=\textwidth]{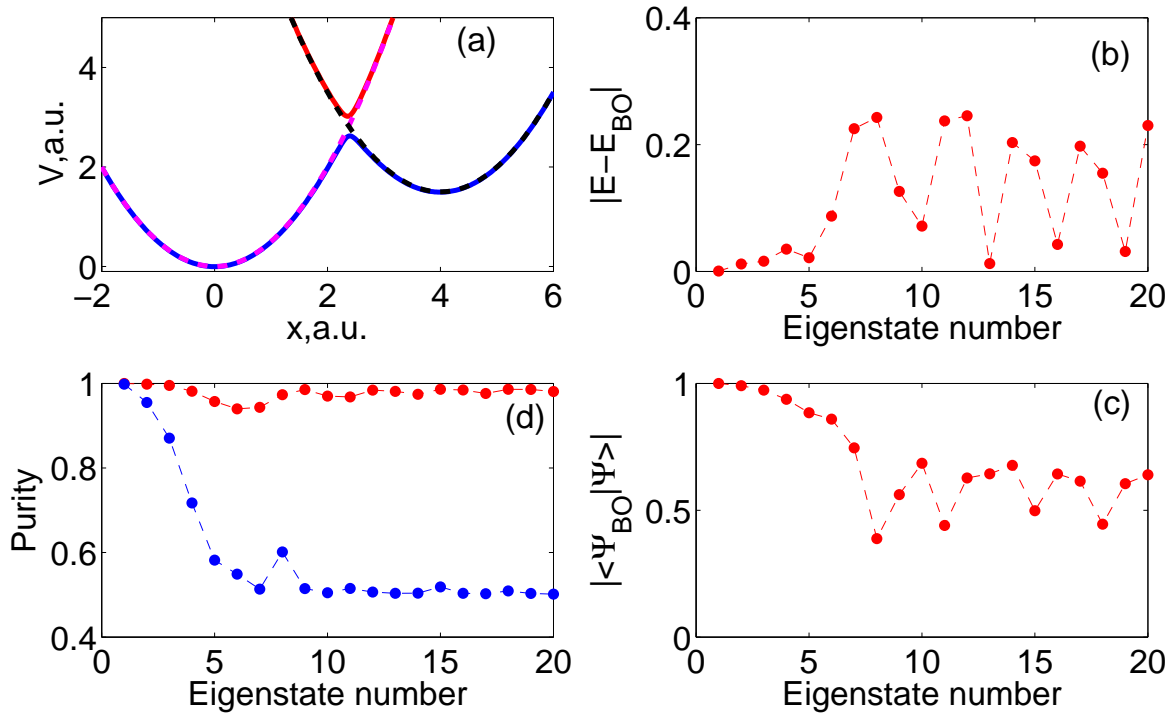}
  \caption{Model 2: (a) Adiabats (solid) and diabats (dashed); 
  (b) Absolute energy differences between exact and BO eigen-states;
  (c) Absolute overlaps between exact and BO eigen-functions; (d) Purities of the BO (blue) and exact (red) eigen-states.}
  \label{fig:m2}
\end{figure*}

\paragraph{Model 2:} By introducing an  electronic energy shift $\Delta=1.5\omega$ which 
breaks the diabatic vibrational level alignment of Model 1, the exact wave-functions now acquire a high degree of localization.  In perturbation theory terms, 
a relatively small coupling $c=\omega/5$ cannot generate appreciable contributions from the vibrational states of the two diabats when the minimal energy 
difference between levels is $\omega/2$. In turn, in  the BO approximation 
this localization is present only in the first few  states which are localized in the lower energy well. 
Delocalization of the higher energy states in BO leads to a failure of the BO approximation for 
these states (\figs{fig:m2}b and c).     The localization in the exact wave-functions leads to $ S_{12}\approx 0$ and 
$ S_{11} S_{22}\approx 0$ in \eq{eq:pur2s}, and therefore, 
the purity for all states is very close to 1 (\fig{fig:m2}d).
In contrast, the purity in the BO approximation quickly drops to 
$1/2$ because of the states' delocalization [\eq{eq:deloc}].  

This model exemplifies a case in which the exact eigenstates are approximately unentangled, while the corresponding BO states can be strongly entangled. For some of these states (i.e. state 5) the disparity between the degree of entanglement between the exact and BO states leads to only modest energetic errors. That is, the BO state can be adequate from an energetic perspective even when the entanglement content of the BO state is a poor approximation to the exact eigenstate.

\begin{figure*}
  \centering
  \includegraphics[width=\textwidth]{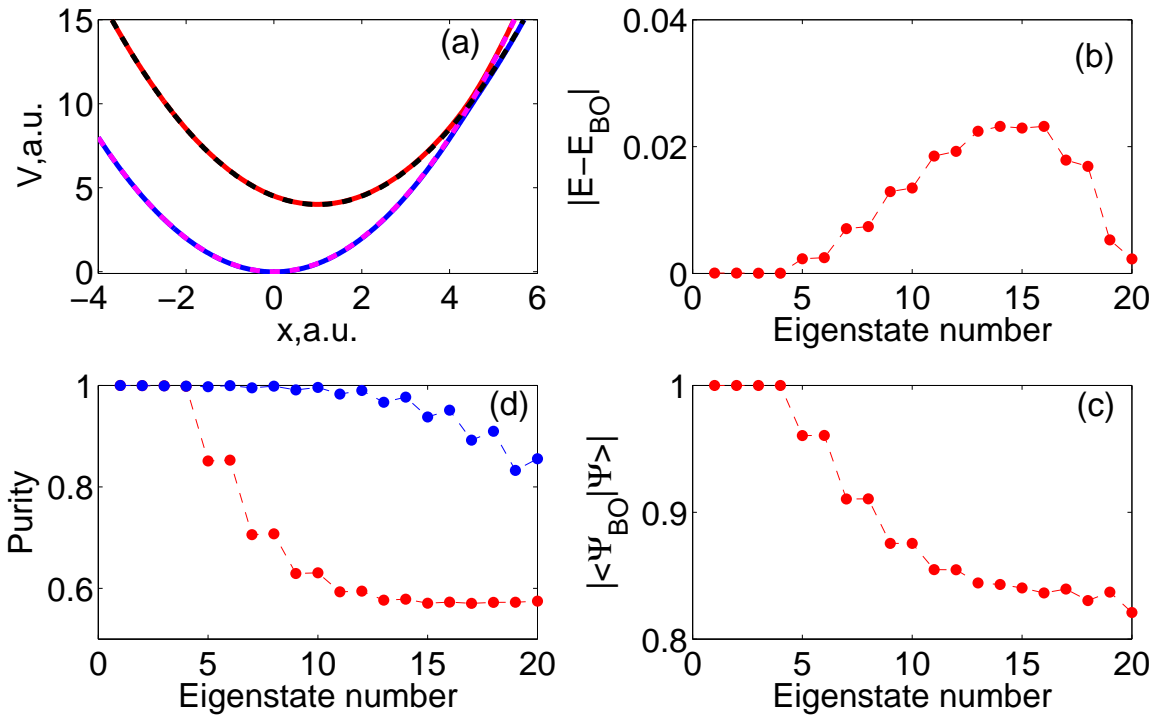}
  \caption{Model 3: (a) Adiabats (solid) and diabats (dashed); 
  (b) Absolute energy differences between exact and BO eigen-states;
  (c) Absolute overlaps between exact and BO eigen-functions; (d) Purities of the BO (blue) and exact (red) eigen-states.}
  \label{fig:m3}
\end{figure*}

\paragraph{Model 3:} Owing to the electronic energy shift $\Delta=4\omega$ that 
preserves an energetic alignment of {the} diabatic vibrational levels, and a smaller coordinate 
shift, $a=1$, {the} high energy exact wave-functions {of this model} consist of almost equal  
contributions from the two diabatic states. 
Nuclear functions $\tilde{\chi}_1$ and $\tilde{\chi}_2$ corresponding to these contributions 
 are almost orthogonal ($S_{12}\approx 0$ in \eq{eq:pur2s}) 
due to a very different number of nodes in energetically aligned vibrational states from the two 
parabolas. 
This leads to the purity close to $1/2$ for these states (\fig{fig:m3}d).

As in Model 2, the BO approximation is adequate only for 
lower states where both methods produce localized nuclear wave-functions (\figs{fig:m3}b and c). In this case, the purity and energy of the BO states is an excellent approximation to the exact states. 
Naturally, the purity for these low states is close to 1 (\fig{fig:m3}d). 
However, in contrast to Model 2, 
the purity of BO states stays close to 1 even for higher excited states because of persistent 
localization of states that makes the product $S_{11}S_{22}$  and overlaps  $S_{12}$ in 
\eq{eq:pur2s} small.  
In spite of the higher value of $\gamma$ in this model wth respect to the other models, 
localized states occur here 
because of a disproportional partitioning of the nuclear BO wave-function into large and small 
(norm-wise) diabatic components $\tilde{\chi}_1$ and $\tilde{\chi}_2$. 
This disproportionality originates from 
the right shift of the diabatic intersection point, $b=4$ (\fig{fig:m3}a), which regulates 
the partitioning location (see \fig{fig:part}). Such a right shifted
partitioning of the BO nuclear probability density produces one nuclear component which is dominant 
and the other one that only represents a small tail of the original distribution.
Thus, in Model 3, we have states where the BO approximation breaks down even when the BO eigenstates are only weakly entangled.

\subsubsection {{Entanglement in non-stationary states}}

 \begin{figure}
  \includegraphics[width=0.49\textwidth]{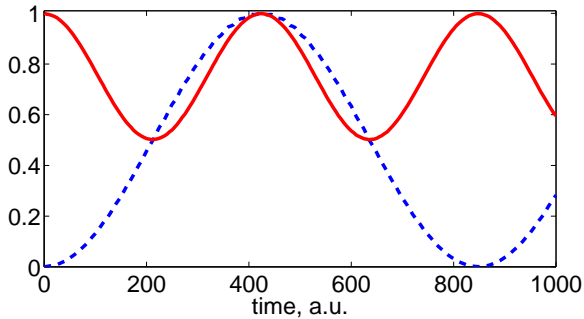}
  \caption{The average nuclear position (dashed blue) and purity $P_N$ (solid red) 
 as functions of time for Model 1 BO dynamics with a non-stationary nuclear wave-packet.}
  \label{fig:m1dyn}
\end{figure}
  \begin{figure}
  \includegraphics[width=0.49\textwidth]{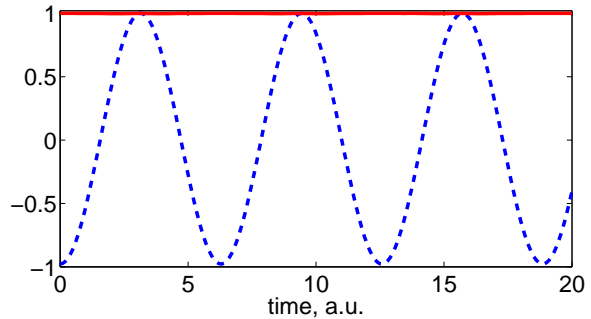}
  \caption{The average nuclear position (dashed blue) and purity $P_N$ (solid red) 
 as functions of time for Model 3 BO dynamics with a non-stationary nuclear wave-packet.}
  \label{fig:m3dyn}
\end{figure}

To illustrate the behavior of the purity for non-stationary BO states we now  consider two cases where the
dynamics is adequately represented by the BO approximation.  
First, in Model 1, an initial wave-packet has been taken as a ground state of the diabatic 
uncoupled parabola
\begin{equation}
  \label{eq:Gaussian_wp}
  \chi(x) = \left(\frac{\omega}{\pi}\right)^{1/4}
  \exp{\left(-\frac{\omega(x-x_0)^2}{2} \right)},
\end{equation} 
it has been centred at the bottom of the left well $x_0=0$. 
In the BO representation, this wave-packet is mostly comprised of the two lowest 
energy eigen-states of the double well problem; symmetric and anti-symmetric wave-functions
which after summation give localization in a single well.
Due to the superposition nature, this initial nuclear wave-packet 
will tunnel back and forth between two wells. 
 Figure~\ref{fig:m1dyn} presents both coherent oscillations of the left 
 well population and the purity dynamics. Naturally, the purity is $1$ at the end points corresponding 
 to localization of a wave-packet within a particular well, while it drops as low as 
 $1/2$ during the period of coherent oscillations. 
 
 Second, in Model 3, the same initial wave-packet [\eq{eq:Gaussian_wp}] has been 
 placed on the left slope ($x_0=-1$ a.u.) 
 of the lower potential. This wave-packet represents a coherent superposition of low energy BO 
 vibrational states  which do not have enough energy to transfer on to the upper BO electronic state. 
 Thus, wave-packet dynamics represents vibrational coherent oscillations at the bottom of the lowest BO state 
without changing the purity of the wave-packet over time (\fig{fig:m3dyn}). 

\subsubsection{Criterion for disentanglement of BO states}

For devising a simple qualitative picture to understand entanglement for an arbitrary  
electron-nuclear state, it is useful to introduce the notion of nuclear function support.
We define a function support as a collection of $x$-ranges where the function has 
non-negligible value. Then we introduce a domain of adequacy for each CBO 
configuration as an $x$-range where the BO electronic wave function has a dominant 
contribution from this CBO configuration (e.g., $|C_i(\vect{R})|\gg |C_j(\vect{R})|$ for $\forall j\ne i$). 
There are also intermediate regions in the $x$-space 
where the BO electronic function can have comparable contributions from different 
CBO configurations.
If the nuclear component has a functional support in these regions 
the BO approximation can become inadequate, and therefore, 
entanglement considerations would require accounting for nonadiabatic effects.   
In cases where the BO approximation is adequate and the nuclear function support
is located only in domains of adequacy for single CBO configurations, 
a simple estimate can be made for the BO state purity. If the support of 
the nuclear function $\chi (x)$ spans $N_D$ 
single CBO configuration domains, $\mathcal{D}_i$, then the purity will be 
\bea
P_N\approx\sum_{i=1}^{N_D} \omega_i^2, 
\eea
where individual domain weights are given by 
\bea
\omega_i =  \int_{\mathcal{D}_i} |\chi(x)|^2 dx.
\eea
Thus, if we extend our model to $N$ 1D parabolic potentials  
all shifted along the $x$-axis consequentially from the origin and constantly coupled, then 
the purity of the ground state will be $1/N$. This setup can be thought as
a finite model for a periodic system, and it shows that entanglement of the ground BO state 
can be made indefinitely strong by increasing $N$. 

\section{Final Remarks}
\label{sec:concl}

The formal and numerical results presented above show that when the BO approximation is exact  ($\gamma\rightarrow\infty$ for the AC model) the resulting BO states are unentangled. However, in the usual situation in which the BO strategy is an approximation the resulting BO states will generally be entangled.  Contrary to intuition, we find that while non-adiabatic couplings can lead to electron-nuclear entanglement and to a failure of the BO approximation, the degree of entanglement of a BO state and the degree of validity of the BO approximation are generally uncorrelated.  Thus, it is possible to find accurate BO states with a high 
degree of entanglement and poor BO states with a low entanglement level. Further, the purity of the BO states can either be 
higher or lower than that of the exact eigenstates.

The reason for this counterintuitive behavior is that while the degree of entanglement of BO states is determined by their deviation from the corresponding states in  the crude BO approximation, the accuracy of the BO approximation is dictated, instead, by the deviation of the BO states from the exact electron-nuclear states. These two metrics are not necessarily simply connected and this explains the absence of an apparent correlation. 

The intuitive picture is restored in the limit where the BO states   coincide with the CBO states (c.f. the first few levels in Model 3). In this limit, any  entanglement in the BO state  also signals a decay in the validity of the BO approximation.  By contrast, when the BO states are very different from the corresponding CBO states, this intuition does not hold any more. This was dramatically illustrated by the double-well problem in Model 1, that involve ``nonlocal'' BO states with nuclear probability amplitude associated with  distinct electronic \emph{diabatic} states. These states are seen to be strongly entangled even when the BO states provide a useful approximation to the exact states. 

In fact, we find that a more adequate criterion for unentanglement of BO states is to require  that the nuclear wave-function support is within the domain of adequacy of a single CBO configuration. Entanglement in the BO state is inevitable  if this support spans a region where more than one CBO 
configuration contributes to the BO state.   The implication is that in molecules, electron-nuclear entanglement 
 and thus electronic decoherence can occur even in the ground-state, zero-temperature, BO approximation. 
 
Importantly, we observe that the BO approximation does not necessarily preserve the entanglement character of the exact states even when the BO approximation is adequate from an energetic perspective. This fact  complicates the interpretation of coherence phenomena for electrons in molecules. This is because the degree of coherence for the electronic subsystem when the electron-nuclear system is in a given superposition of exact eigenstates can be very different from that predicted by the same superposition but among the BO equivalents to the exact eigenstates.   This implies that analyzing electronic coherences starting from BO states for the system plus bath should involve consideration of how accurately those states preserve the entanglement properties of the exact states.

\section{Acknowledgements}

A.F.I. thanks R. Kapral, I. G. Ryabinkin, and P. Brumer for helpful discussions and 
acknowledges funding from the Natural Sciences and Engineering 
Research Council of Canada (NSERC) through the Discovery Grants Program 
and the Alfred P. Sloan Foundation. I.F. acknowledges support by the 
National Science Foundation under CHE - 1553939. 

\bibliography{BO_entang.bbl}

\end{document}